\documentclass[prb,aps,twocolumn,showpacs]{revtex4}

\usepackage{graphicx}

\begin{document}

\title{Spin Fluctuations and Frustrated Magnetism in the Multiferroic FeVO$_4$}

\author{J. Zhang$^1$}
\author{L. Ma$^2$}
\author{J. Dai$^2$}
\author{Y. P. Zhang$^2$}
\author{Z. Z. He$^3$}
\author{B. Normand$^2$}
\author{Weiqiang Yu$^2$}

\address{$^1$School of Energy, Power and Mechanical Engineering, North 
China Electric Power University, Beijing 102206, China}
\address{$^2$Department of Physics, Renmin University of China, Beijing 
100872, China}
\address{$^3$State Key Laboratory of Structural Chemistry, Fujian Institute 
of Research on the Structure of Matter, Chinese Academy of Sciences, Fuzhou, 
Fujian 350002, China}

\date{\today}

\begin{abstract}

We report $^{51}$V nuclear magnetic resonance (NMR) studies on single crystals 
of the multiferroic material FeVO$_4$. The high-temperature Knight shift shows 
Curie-Weiss behavior, $^{51}K = a/(T + \theta)$, with a large Weiss constant 
$\theta \approx$ 116 K. However, the $^{51}$V spectrum shows no ordering near 
these temperatures, splitting instead into two peaks below 65 K, which 
suggests only short-ranged magnetic order on the NMR time scale. Two magnetic 
transitions are identified from peaks in the spin-lattice relaxation rate, 
$1/^{51}T_1$, at temperatures $T_{N1} \approx$ 19 K and $T_{N2} \approx$ 13 K, 
which are lower than the estimates obtained from polycrystalline samples.  
In the low-temperature incommensurate spiral state, the maximum ordered 
moment is estimated as 1.95${\mu}_B$/Fe, or 1/3 of the local moment. Strong 
low-energy spin fluctuations are also indicated by the unconventional
power-law temperature dependence $1/^{51}T_1 \propto T^2$. The large Weiss 
constant, short-range magnetic correlations far above $T_{N1}$, small ordered 
moment, significant low-energy spin fluctuations, and incommensurate ordered 
phases all provide explicit evidence for strong magnetic frustration in 
FeVO$_4$.

\end{abstract}

\maketitle

\section{Introduction}

In multiferroic materials such as $R$MnO$_3$ (R: rare earths) 
\cite{Kimura_Nature_2003}, Ni$_3$V$_2$O$_8$\cite{Lawes_PRL_2005}, and 
MnWO$_4$\cite{Taniguchi_PRL_2006}, ferroelectricity is believed to be 
driven by magnetism because it emerges simultaneously with a spiral 
magnetic order upon cooling. These materials are known as ``type-II'' 
multiferroics,\cite{Khomskii_Physics_2009} and their ferroelectricity 
as ``improper.''\cite{Cheong_NM_2007} Even though such multiferroicity 
has to date been found only at low temperatures, it has attracted very 
significant research interest due to the possibilities it offers for 
tunable multiferroic devices. Phenomenologically, the breaking of 
magnetic inversion symmetry and strong Dzyaloshinskii-Moriya (DM) 
interactions are believed to be essential ingredients for understanding 
the coupling between ferroelectricity and magnetism.\cite{Cheong_NM_2007} 
Theoretically, inversion symmetry-breaking is thought to be caused by 
strong magnetic frustration. Experimentally, while magnetic frustration 
is certainly suggested by the available susceptibility 
data,\cite{He_JSSC_2008} the coupling between ferroelectricity 
and magnetism has been difficult to quantify, and more spectroscopic 
studies are required to confirm its existence and explore its origin.

\begin{figure}
\includegraphics[width=7.5cm]{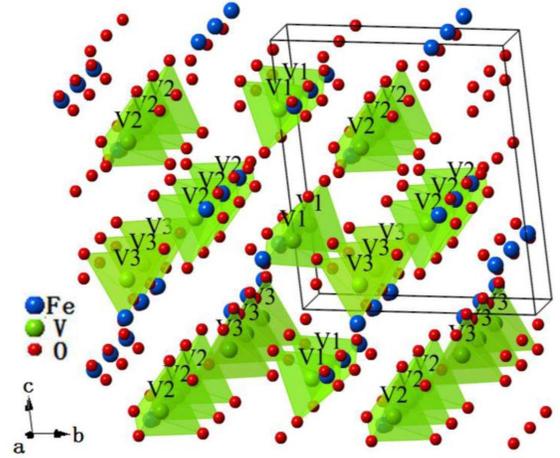}
\caption{\label{struc1}(color online) Crystal structure (triclinic) of 
FeVO$_4$ viewed along the $a$ axis. The three different V sites are 
indicated, each occurring as a VO$_4$ tetrahedron.}
\end{figure}

FeVO$_4$ is a multiferroic compound that has been characterized mostly 
in polycrystalline\cite{Aladine_PRB_2009, Kundys_PRB_2009, MuMK_SSC_2011}
or thin-film form,\cite{Dixit_JPCM_2009, Dixit_PRB_2010} although single 
crystals have also been synthesized.\cite{He_JSSC_2008} It crystallizes 
in a triclinic structure, with each unit cell containing 6 Fe$^{3+}$, 
6 V$^{5+}$, and 24 O$^{2-}$ ions, as shown in Fig.~\ref{struc1}. Neutron 
scattering studies of polycrystal powders\cite{Aladine_PRB_2009} indicate 
a collinear incommensurate antiferromagnetic (AFM) order below $T_{N1} 
\approx$ 22 K and a non-collinear incommensurate, or spiral, AFM order 
below $T_{N2}\approx$ 15 K. Ferroelectricity occurs only below $T_{N2}$, a 
result found also in other type-II multiferroic materials.\cite{Cheong_NM_2007} 
To date, however, most studies of FeVO$_4$ have focused on this 
ferroelectricity,\cite{Kundys_PRB_2009, MuMK_SSC_2011, Dixit_JPCM_2009, 
Dixit_PRB_2010} and there has been only limited investigation of its 
magnetic states and their relation to frustration. The availability of 
FeVO$_4$ as high-quality single crystals makes it an excellent target 
material for NMR studies of its magnetic properties and of their coupling 
with ferroelectricity.

NMR is a low-energy, local probe, which is in principle an ideal tool for 
the study of multiferroics. It is sensitive both to ferroelectricity, through 
the coupling between the nuclear quadrupole moment and the electric field 
gradient (EFG), and to magnetism, through the hyperfine coupling between 
the nuclei and the magnetic moments. In FeVO$_4$, however, only the $^{51}$V 
signal could be observed within our available NMR window and the weak 
quadrupole moment ($-0.05$ barns) of this nucleus makes the coupling to the 
EFG too weak to be detected. Therefore we focus primarily on the low-energy 
magnetic properties, which reveal strong magnetic frustration in this system, 
and thus also provide essential information for understanding the nature of 
multiferroicity.

In this paper we present all of the information about magnetism in FeVO$_4$ 
that can be deduced from $^{51}$V NMR. The high-temperature spectrum is 
single-peaked in the paramagnetic phase, where the Knight shift, like the 
susceptibility, follows a Curie-Weiss form with a large Weiss constant. On 
cooling below 65 K, the spectrum splits into two peaks, which indicates 
a local symmetry-breaking effect on the NMR time scale, or the onset of 
short-range magnetic order far above the magnetic transition. From the 
spin-lattice relaxation time we identify the two magnetic transitions at 
$T_{N1}$ = 19 K and $T_{N2} =$ 13 K, consistent with other results reported 
on single crystals, but lower than those for powder samples. Below $T_{N2}$, 
we find a spectrum with no applied field, which reflects a complicated 
distribution of hyperfine fields transferred from the incommensurate magnetic 
moments of the Fe ions. However, the upper bound for the ordered moment is 
only 1.95${\mu}_B$/Fe, far less than the paramagnetic local moment, while 
the temperature dependence of $1/^{51}T_1$ below $T_{N2}$ suggests strong 
low-energy spin fluctuation effects. We discuss how all of these features  
are the fingerprints of strongly competing magnetic interactions. 

The structure of our report is as follows. In Sec.~II we describe briefly 
the NMR methods as applied to FeVO$_4$. In Sec.~III we present all of the 
results we obtain for $^{51}$V spectra over a range of temperatures, for 
the relaxation times $1/T_1$ and $1/T_2$, for the Knight shift $^{51}K$, and 
at zero magnetic field. In Sec.~IV we discuss the implications of our results 
for frustrated magnetism in FeVO$_4$ and it connection to multiferroicity. 
Section V contains a brief summary. 

\section{Material and Methods}

Single crystals of FeVO$_4$ were synthesized by the flux-growth method, 
using V$_2$O$_5$ as the self-flux.\cite{He_JSSC_2008} The crystal structure 
is shown in Fig.~\ref{struc1} and the crystal alignment, required for NMR, 
was determined by von Laue measurements. The complete $^{51}$V NMR spectra 
were measured by field sweeps of the spin-echo signal, using a fixed 
frequency of 111.7 MHz and sweeping the magnetic field with its orientation 
parallel to the (0 1 -1) plane of the crystal. The spin-echo pulse sequence 
was $\pi/2-\tau-\pi$, with $\tau = 10$ $\mu$s and respective $\pi/2$ and 
$\pi$ pulse lengths of 2.5 and 4 $\mu$s. The Knight shift was determined 
in the paramagnetic state from $^{51}K = (f - {^{51}\gamma B})/{^{51}\gamma 
B}$, where $f$ is the resonance frequency, $B$ is the applied field, and 
$^{51}\gamma$ = 11.197 MHz/T is the $^{51}$V gyromagnetic ratio. The 
spin-lattice relaxation rate $^{51}T_1$ was measured by the spin-inversion 
method, and the spin-spin relaxation rate $^{51}T_2$ from the spin echoes, 
both of which were found to follow standard, single-exponential spin 
recovery/decay functions.

\begin{figure}
\includegraphics[width=7.5cm]{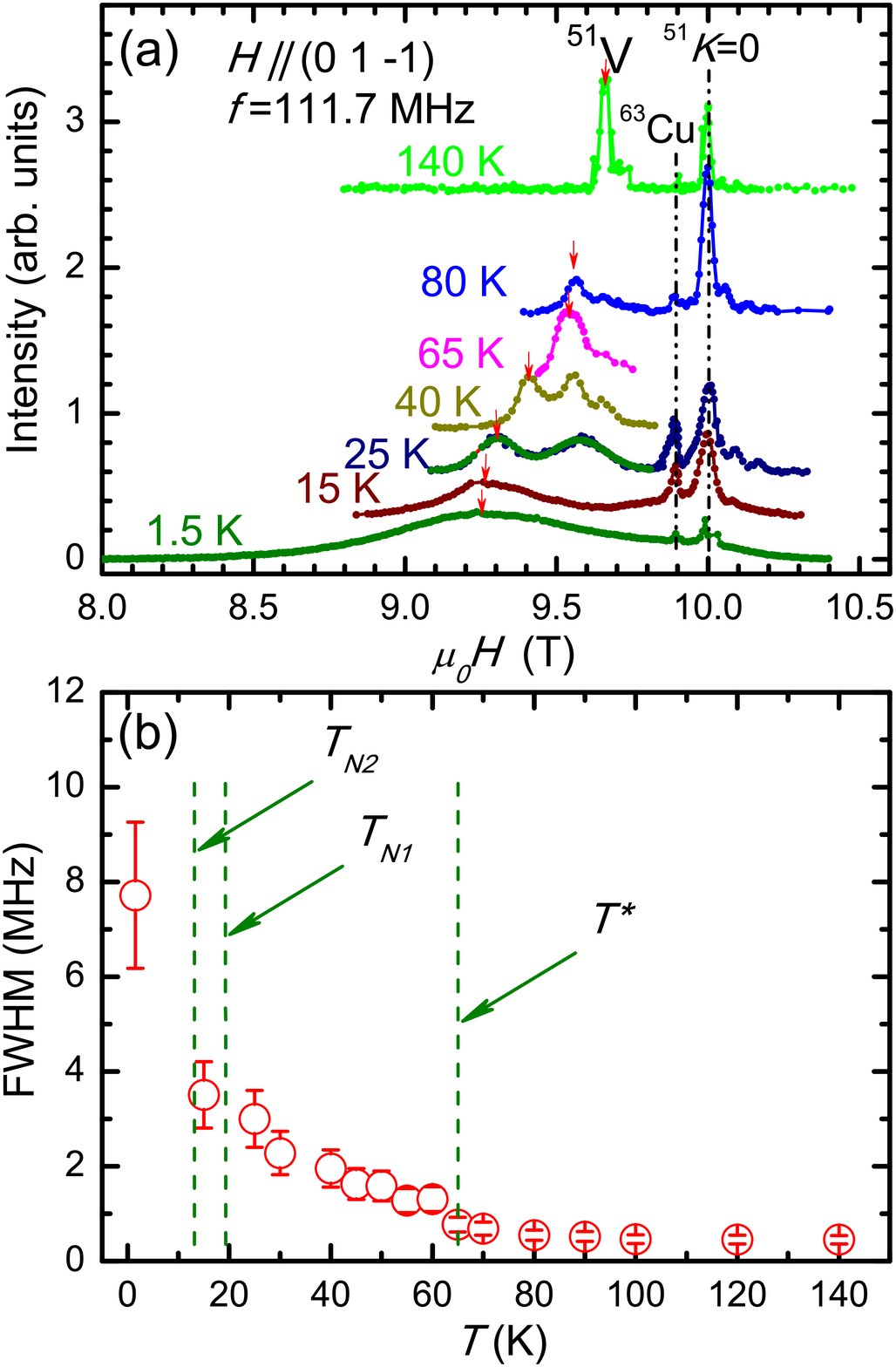}
\caption{\label{spec2}(color online) (a) Field-sweep $^{51}$V spectra over 
the full range of temperatures probed, with the field applied parallel 
to the crystal (0 1 -1) planes. Red arrows indicate the fields used for the 
$1/^{51}T_1$ and $1/^{51}T_2$ measurements shown in Fig.~3. The solid line at 
25 K is a double-Gaussian fit to the spectrum. (b) FWHM of the NMR spectra 
measured under the same field-sweep conditions. The temperatures $T^*$, 
$T_{N1}$, and $T_{N2}$ are introduced in the text. Vertical bars on the data 
points indicate the standard deviations of the fitting curves.}
\end{figure}

Two comments are in order on the $^{51}$V spectrum at $T$ = 140 K, shown in 
Fig.~\ref{spec2}(a), which has a finite Knight shift $^{51}K \sim 3\%$. First, 
from the triclinic crystal structure shown in Fig.~\ref{struc1}, three $^{51}$V 
resonance peaks are expected. Each unit cell in FeVO$_4$ contains three pairs 
of non-identical V$^{5+}$ sites, each linked by lattice inversion symmetry and 
labeled by V1, V2, and V3. All three types of site have different bond lengths 
to all of their neighboring O$^{2-}$ and Fe$^{3+}$ ions, and therefore should 
appear as separate peaks. However, in our data only one of the three site 
pairs is detected within the NMR window. As we will show in Sec.~III, the 
spin-spin relaxation time $T_2$ of our signal is very short (less than 50 
$\mu$s) due to the dominance of magnetic fluctuations. Thus we believe our 
inability to observe the other two $^{51}$V pairs arises because these sites 
have a stronger hyperfine coupling and therefore even shorter $T_2$ times, 
which move the corresponding signals out of our measurement window. Second, 
we did not find measurable quadrupolar effects on the resonance frequency, 
which is reasonable for the center site of the local VO$_4$ tetrahedra and 
for the low quadrupole moment of $^{51}$V. As noted in Sec.~I, the absence 
of significant quadrupole effects prevents us from studying the lattice 
structure, and therefore the ferroelectricity, directly. 

\section{NMR Measurements}

We begin our investigation of the NMR response of FeVO$_4$ single crystals 
by considering the high-field $^{51}$V spectra at elevated temperatures. 
Spectra starting at 140 K are shown in Fig.~\ref{spec2}(a). Between 140 K 
and 80 K they show a single peak at a field of 9.66 T, which is a regular 
paramagnetic signal. As we decrease the temperature to 1.5 K, it is clear 
that the line shape, line width, and peak frequency change dramatically, 
reflecting a rich variety of magnetic properties for a single material. 
We note that a spurious $^{51}$V signal is also observed at 10.0 T ($^{51}K
 = 0$), which has a $T_1$ value more than three orders of magnitude higher 
than the other peaks and does not sense the magnetic transitions on cooling, 
and thus is probably the consequence of a weak impurity phase or an inclusion 
of the crystal-growth flux. A weak and asymmetric shoulder feature at 9.65 T 
is also visible at temperatures between 40 and 80 K before being lost as 
the spectrum broadens, and this may reflect some dilute local disorder.  

Figure \ref{spec2}(a) shows representative spectra in all of the different 
magnetic phases of FeVO$_4$. Below a temperature $T^* = 65$ K, the $^{51}$V 
spectrum broadens and develops a prominent double-peak feature suggestive 
of strong spin correlations, or short-range magnetic order on the NMR time 
scale; we discuss this interpretation in detail below. The spectra at 15 K 
and at 1.5 K show such significant broadening that they again appear to 
have a single peak. This form is typical for systems with spatially
inhomogeneous magnetic order when subject to an applied field, because the 
nuclei respond to both the external field and the varying internal field 
in the crystal. We recall here (Sec.~I) that at 15 K the system is in the 
collinear incommensurate AFM phase while at 1.5 K it is in the incommensurate 
spiral AFM phase. The full width at half-maximum height (FWHM) of the peak, 
shown in Fig.~\ref{spec2}(b), quantifies the increase in broadening with 
cooling as the sample steps through the sequential changes of magnetic 
properties.

\begin{figure}
\includegraphics[width=8cm]{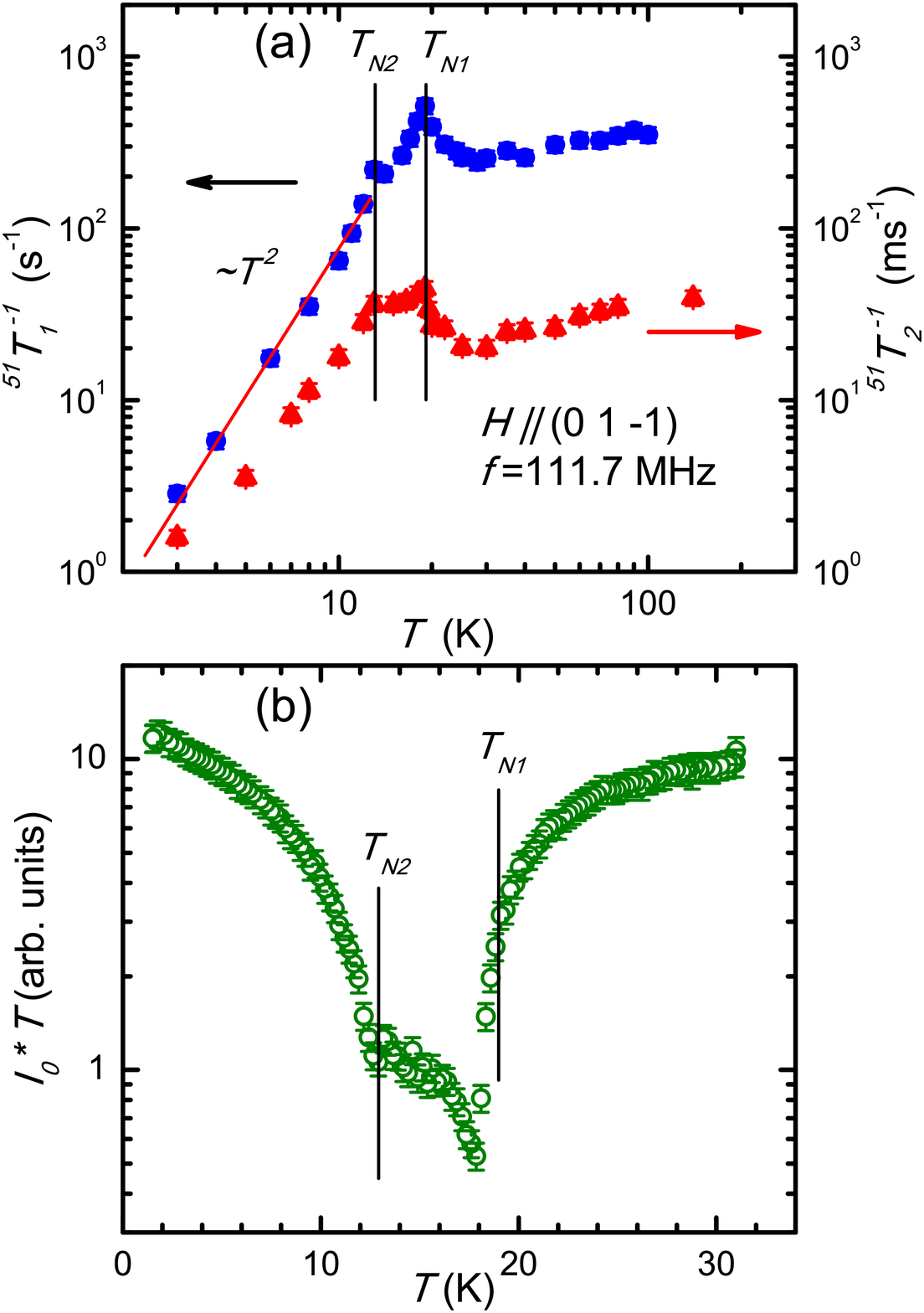}
\caption{\label{t1vst3}(color online) (a) Spin-lattice relaxation rate 
$1/^{51}T_1$ and spin-spin relaxation rate $1/^{51}T_2$ as functions of 
temperature, measured at the peak position of the spectra. (b) Temperature 
dependence of the Boltzmann-normalized echo intensity at the peak position 
of the $^{51}$V spectrum.}
\end{figure}

To identify the magnetic transition temperatures, we measured the 
spin-lattice relaxation rate $1/^{51}T_1$ and the spin-spin relaxation 
rate $1/^{51}T_2$. We show results for both quantities, measured for each 
temperature at the peak positions on the low-field side of the spectra 
[red arrows in Fig.~\ref{spec2}(a)]; below $T_{N2}$, $1/^{51}T_1$ is lowest 
at the peak frequency and increases by approximately $30\%$ across the 
frequency range. The temperature dependence of $1/^{51}T_1$, shown in 
Fig.~\ref{t1vst3}(a), first displays a slow decrease with temperature 
upon cooling below 100 K. However, there is a prominent increase below 30 K 
leading to a peak at 19 K, which indicates the first magnetic phase transition 
at $T_{N1}$. Upon further cooling, $1/^{51}T_1$ decreases again, although a 
small second peak is resolvable at $T = 13$ K, indicating the second magnetic 
transition ($T_{N2}$). Both transitions can also be resolved in the $1/^{51}T_2$ 
data, also shown in Fig.~\ref{t1vst3}(a), where $1/^{51}T_2$ exhibits the same 
behavior as $1/^{51}T_1$; two peaks in the relaxation rate are clearly formed 
at the two magnetic transition temperatures $T_{N1} = 19$ K and $T_{N2} = 13$ K. 
While $1/T_1$ in a magnetic system is expected to be controlled by spin 
fluctuations, this is in general not so clear for $1/T_2$. However, the fact 
that $1/^{51}T_2$ peaks at the transitions and shows a temperature dependence 
so similar to that of $1/^{51}T_1$ indicates that $1/^{51}T_2$ is indeed 
dominated by the magnetic fluctuations in FeVO$_4$. We comment again here 
that the $^{51}T_2$ values shown in Fig.~\ref{t1vst3}(a) are very short (under 
50 $\mu$s over much of the temperature range), a result we ascribe to these 
very strong magnetic fluctuations and which we believe prevents our detection 
of the other two V site pairs (Sec.~II). 

To verify the nature of both transitions we have also measured the echo 
intensity $I_0$ at the peak frequency, integrated over a finite frequency 
range. The Boltzmann-corrected echo intensity, $I_0T$, is shown in 
Fig.~\ref{t1vst3}(b) and is inversely proportional to both FWHM and 
$e^{2\tau/T_2}$, where $\tau$ is the spin-echo refocusing time. As the 
temperature is lowered from 30 K to $T_{N1}$, $I_0T$ decreases strongly 
due to the combination of inhomogeneous line-width broadening 
[Fig.~\ref{spec2}(b)] and the temperature dependence of $T_{2}$
[Fig.~\ref{t1vst3}(a)]. In fact $^{51}T_2 \sim 20 \mu$s is very short at 
$T_{N1}$ [Fig.~\ref{t1vst3}(a)] because of the strong magnetic fluctuations, 
and this causes a large signal loss in our spin-echo measurements. The 
rise in echo intensity below $T_{N1}$ is caused by the fall in $1/^{51}T_2$ 
[Fig.~\ref{t1vst3}(a)]. However, the increase below $T_{N2}$
[Fig.~\ref{t1vst3}(b)] occurs despite the strong spectral broadening at 
these temperatures [Fig.~\ref{spec2}(b)] and cannot be explained by $T_2$ 
effects alone. While the kink directly below $T_{N2}$ is caused by the sharp 
fall in $1/^{51}T_2$ [Fig.~\ref{t1vst3}(a)], the continued increase in echo 
intensity at low temperatures, once $T_2$ is very long again, is due to RF 
enhancement of the NMR signal. This phenomenon is typical in ordered 
magnetic systems and has also been reported in multiferroic materials such 
as TbMn$_2$O$_5$;\cite{Baek_PRB_2006} in this compound it shows a hysteresis 
effect at high fields that the authors proposed may originate in a coupling 
between the AFM domain walls and ferroelectric domain walls. In FeVO$_4$ we 
do not find evidence for hysteresis effects, and we suggest that the RF 
enhancement below $T_{N2}$ is intrinsic in a spiral magnet.
 
\begin{figure}
\includegraphics[width=8cm]{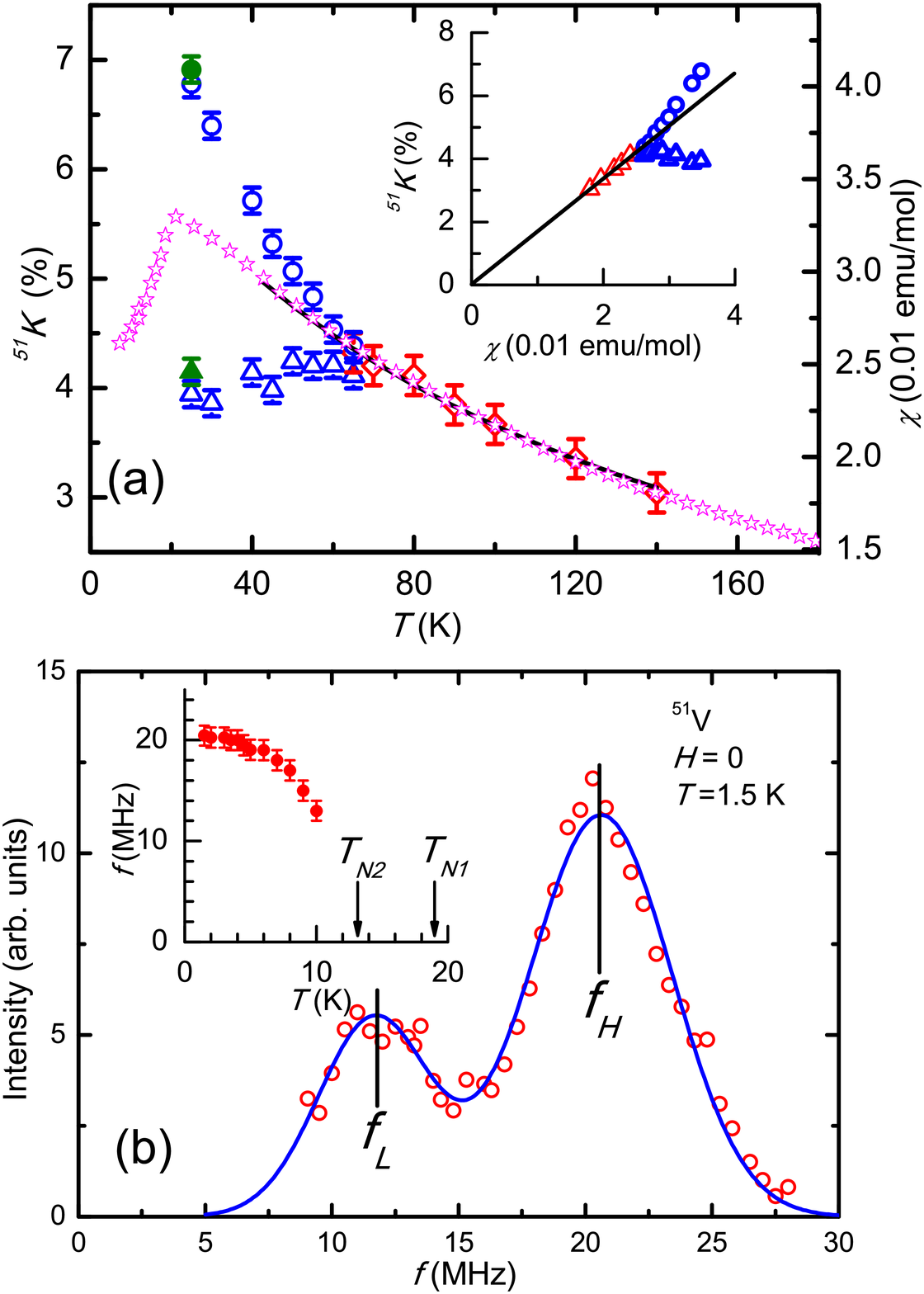}
\caption{\label{knvst4}(color online) (a) Temperature dependence of the 
Knight shift $^{51}K$ (left axis) measured at the peak frequency of the 
$^{51}$V spectrum in the paramagnetic phase. Knight-shift data shown as 
open diamond, circle, and triangle symbols are measured under a field of 
9.5 T, data shown as solid symbols at 4 T. The bulk susceptibility (star 
symbols, right axis, adapted from Ref.~\onlinecite{He_JSSC_2008}) is shown 
for comparison. Inset: $^{51}K$ as a function of the dc susceptibility with 
temperature as the implicit parameter. Solid lines in both panels are fits 
to the Knight-shift data using the Curie-Weiss function. (b) Zero-field 
$^{51}$V NMR spectrum measured at $T$ = 1.5 K. $f_H$ and $f_L$ label the 
resonance frequencies of the high- and low-frequency peaks. Inset: $T$ 
dependence of $f_H$. The arrows indicate the two magnetic transition 
temperatures.}
\end{figure}

\section{Magnetic Properties}

In this section we discuss the distinctive magnetic properties of FeVO$_4$ 
revealed by NMR in the different regimes of temperature. We focus first on 
the properties in the truly paramagnetic phase above $T^* = 65$ K. Figure 
\ref{knvst4}(a) shows the Knight shift, $^{51}K$, deduced from the peak 
frequency of the spectrum, as a function of temperature. The Knight shift 
measures the spin susceptibility of the system, and so the monotonic increase 
of $^{51}K$ on cooling is consistent with paramagnetic behavior. In fact 
$^{51}K(T)$ can be fitted very well by a Curie-Weiss form, $^{51}K = A/(T + 
\theta)$, with the Weiss constant $\theta \approx 116\pm 15$ K; the fit is 
shown in Fig.~\ref{knvst4}(a). In the inset we show $^{51}K$ against the dc 
susceptibility (adapted from Ref.~\onlinecite{He_JSSC_2008}) with temperature 
as the implicit parameter, from which we estimate the hyperfine coupling 
constant to be $^{51}A_{\rm hf} \approx 9.37 \pm 0.23$ kOe/$\mu_B$. 

Values of $\theta$ reported in the literature show some considerable variation, 
with Curie-Weiss fits to susceptibility data from polycrystals giving $\theta 
\approx$ 125 K\cite{Aladine_PRB_2009} and from single-crystal measurements 
giving $\theta \approx$ 97 K.\cite{He_JSSC_2008} This spread of results may 
reflect an important role for domain-wall effects, a topic to which we 
return below. Two incontrovertible statements are that our measurements are 
fully consistent with previous studies and that they are consistent with 
$\theta/T_{N1} \gg 1$. Such a large value of $\theta$ compared to $T_{N1}$ is 
typical for a number of type-II multiferroic systems,\cite{Cheong_NM_2007} 
and is a key piece of evidence for strong magnetic frustration.

We turn next to the double-peak feature in the spectrum at temperatures 
between $T^*$ and $T_{N1}$, where the intensities of the two peaks are 
rather similar and can be fitted rather well by a double-Gaussian function
[Fig.~\ref{spec2}(a)]. The Knight shifts calculated from both peaks are shown 
in Fig.~\ref{knvst4}(a), where it is clear that $T^*$ represents a bifurcation 
in behavior. $^{51}K(T)$ in this temperature range deviates both from the 
high-$T$ Curie-Weiss form and from the high-$T$ linear scaling with the 
bulk susceptibility [Fig.~\ref{knvst4}(a)]. As a consequence of this 
line-splitting, the overall line width of the spectral features also 
broadens significantly below 65 K, as shown in Fig.~\ref{spec2}(b). We 
confirmed (data not shown) that the splitting of the peaks is proportional 
to the external field, which indicates a varying local susceptibility rather 
than any static magnetic ordering. Indeed, such a line-splitting is clearly 
a local symmetry-breaking effect, and far the most probable interpretation 
of our data is that the double-peak spectra are caused by strong spin 
correlations, or equivalently short-range magnetic ordering on the time 
scale of NMR. In particular, the hyperfine fields of the two V sites linked 
by lattice inversion symmetry may be different in this short-range-ordered 
state, splitting the spectrum into two peaks with equal intensity as observed.
While the splitting we observe could also be caused by a breaking of crystal 
symmetry, no structural measurements have yet detected such a process at 
temperatures as high as $T^*$. Further evidence in favor of a short-range 
ordering scenario for temperatures $T_{N1} < T < T^*$ in FeVO$_4$ can be 
found by comparison with the situation when $T < T_{N2}$, where the magnetic 
order is long-ranged and static, magnetic inversion symmetry is 
broken,\cite{Aladine_PRB_2009} and our zero-field NMR spectrum also 
resolves a double-peak feature (below).

Strong spin correlations or short-ranged magnetic order above $T_{N1}$ in 
FeVO$_4$ have been also been proposed to interpret specific-heat measurements, 
where a significant absence of magnetic entropy (or an ``entropy recovery'') 
is observed over a broad temperature range above $T_{N1}$.\cite{He_JSSC_2008, 
Aladine_PRB_2009,Dixit_JPCM_2009} While magnetic heat-capacity measurements 
are complicated by issues including phonon subtraction, the line splitting 
and broadening we observe by NMR provide direct evidence for short-range 
magnetic ordering. Our results also give the first accurate measurement of 
the onset temperature $T^*$. Short-range ordering of this type has also 
been reported by NMR, on the basis of inhomogeneous line-width broadening, 
in the materials LiCuVO$_4$\cite{Buttgen_PRB_2010} and 
LaMn(O$_{1-x}$F$_x$)$_3$.\cite{Milhalev} The appearance of short-range magnetic 
order in the paramagnetic state is a first indication of magnetic frustration 
effects, which are necessary to suppress a static AFM order in this regime. 
The fact that $T^*$ is over three times the size of $T_{N1}$, making the 
short-range-ordered region remarkably broad, suggests that frustration is 
very strong in FeVO$_4$. This observation is fully consistent with the 
magnetic frustration revealed by the large Weiss constant in the Knight shift.

The tendency towards short-ranged magnetic order may be an important ingredient 
in explaining the discrepancies between N{\'{e}}el temperatures reported in the 
literature. The values we obtain from NMR, $T_{N1} = 19$ K and $T_{N2} = 13$ K, 
are consistent with the magnetization measurements also performed on single 
crystals.\cite{He_JSSC_2008} However, values reported for powder 
samples\cite{Aladine_PRB_2009,Kundys_PRB_2009} are $T_{N1} = 22$ K and 
$T_{N2} = 15$ K, respectively 3 K and 2 K higher, which represent a 
discrepancy well beyond the expected error bars of the individual 
measurements. We suggest that the transition temperatures in powder samples 
can be enhanced both by grain-boundary and domain-wall effects and by strain 
effects. Given that short-range magnetic ordering already occurs at 65 K, 
both sets of effects provide a ready source of pinning for fluctuating 
magnetic moments, particularly when spiral orientations are favored. Strain 
effects have been found to be very effective in enhancing multiferroic 
properties in a number of compounds.\cite{Lee_PRL_2010} 

Next we discuss the magnetic properties in the ordered phase. In addition to 
our high-field NMR studies, we have also performed zero-field NMR measurements 
to study the magnetic structure in the non-collinear (spiral) phase below 
$T_{N2}$. In Fig.~\ref{knvst4}(b) we show the zero-field $^{51}$V spectrum at 
1.5 K, which has a clear double-peak structure with the two maxima centered 
at $f_H$ (high-frequency) and $f_L$ (low-frequency). The broad spectrum around 
each peak is caused by the distribution of hyperfine fields on the V sites 
transferred from the Fe moments, all of which are different due to their 
incommensurate order (neutron scattering measurements in this phase 
reveal a spiral magnetic modulation period of approximately 100 
nm.\cite{Aladine_PRB_2009}). The FWHM of the high-field spectra at $T = 1.5$ K, 
shown in Fig.~\ref{spec2}(b), is approximately 8 MHz, which is considerably 
less than $f_H$ and therefore indicates that the hyperfine field on the V 
sites is almost perpendicular to the applied external field. The NMR spectrum 
in incommensurate magnetically ordered states usually has a characteristic 
``double-horn'' feature,\cite{Horvatic, Gippius, Pregelj} but this is 
obtained when the applied field is not perpendicular to the internal field. 
Thus in our present field configuration we are not able to distinguish between 
an incommensurate spin structure and other forms of modulation that also 
give rise to a distribution of hyperfine fields, and can state only that our 
broad line shapes are consistent with the known incommensurate order. This 
lack of specificity applies also in the short-range-ordered phase between 
$T_{N1}$ and $T^*$, where we cannot probe the commensurate or incommensurate 
nature of the spin fluctuations. We comment that the incommensurate 
``double-horn'' shape is not similar to the double-peak structures we find 
in either our high-field or zero-field NMR measurements [Figs.~\ref{spec2}(a) 
and \ref{knvst4}(b)]. 

Although the spectral intensity is higher at $f_H$ than at $f_L$, we 
believe this difference is due primarily to the sensitivity of the NMR 
pick-up. This splitting of the spectrum is probably caused by the breaking 
of inversion symmetry in the hyperfine field on the V sites, similar to the 
situation we discussed (on the NMR time scale) in the short-range-ordered 
state, and the default expectation would be peaks of equal weight. We have 
also measured $f_H$ as a function of temperature, finding [inset, 
Fig.~\ref{knvst4}(b)] that it increases significantly on cooling from 
10 K down to 1.5 K, reflecting the development of the ordered moment. By 
using the value of $^{51}A_{\rm hf}$ measured in the paramagnetic phase, the 
high-frequency resonance peak ($f_H$) at zero field sets an upper bound 
for the ordered moment, of 1.95 $\mu_B$/Fe at $T$ = 1.5 K. In the paramagnetic 
phase, however, the magnetization data give a local moment of 5.83 
$\mu_B$/Fe.\cite{He_JSSC_2008} Thus the ordered moment below $T_{N2}$ 
is only 1/3 of the net moment, indicating again the effects of strong 
magnetic frustration even at the lowest temperatures. 

Below $T_{N2}$, the temperature dependence of the spin-lattice relaxation 
rate is different from a conventional antiferromagnet,\cite{Kranendonk_T1} 
where $1/T_1 \sim T^{3}$ due to relaxation by gapless spin waves, and the 
temperature dependence is stronger still in the presence of magnetic 
anisotropy. As shown in Fig.~\ref{t1vst3}(a), $1/^{51}T_1$, measured at 
high field, has power-law behavior below $T_{N2}$ with $1/^{51}T_1 \sim T^{2}$. 
Similar unconventional behavior and anomalously slow spin dynamics have 
been measured in other frustrated magnetic systems, such as 
volborthite,\cite{Yoshida_prl_2009} where they were ascribed to a very 
high density of available low-energy excitations. The low power-law 
temperature dependence found in FeVO$_4$ would seem to indicate the 
presence of persistently strong low-energy spin fluctuations on top 
of the spiral ordered state below $T_{N2}$.

The small ordered moment and the strong low-energy spin fluctuations in 
the spiral magnetic phase reflect once again the effects of magnetic 
frustration in suppressing the ordered moment while enhancing spin 
fluctuations. Combined with the large Weiss constant and the short-range 
magnetic ordering at high temperatures, FeVO$_4$ shows explicit 
evidence of strong frustration all across the phase diagram. Magnetic 
frustration in FeVO$_4$ is clear from the structure shown in Fig.~\ref{struc1}, 
where there are multiple inequivalent Fe--O--Fe and Fe--O--O--Fe paths in 
the system. These paths give rise to effective magnetic coupling processes, 
referred to respectively as superexchange and super-superexchange in the 
structural and magnetic study of Ref.~\onlinecite{Aladine_PRB_2009}, and 
it is reasonable to assume that these interactions compete strongly. Our 
data provide independent evidence reinforcing the presence of strong 
magnetic frustration in FeVO$_4$, and by extension its importance for 
multiferroicity in the form of ferroelectric incommensurate SDW phases. 

The origin of magnetically-driven ferroelectricity in improper multiferroics 
is discussed in Ref.~\onlinecite{Cheong_NM_2007}. Unlike the case of proper 
multiferroics, it does not depend on a ``$d^0$-type'' polar distortion 
mechanism despite the absence of orbital moments\cite{Aladine_PRB_2009} on 
both V$^{5+}$ and Fe$^{3+}$ (which is $d^5$). In fact the breaking of magnetic 
inversion symmetry may in itself not be a sufficient condition, as this is 
broken at $T_{N1}$ in FeVO$_4$, i.e.~in the non-ferroelectric collinear 
incommensurate SDW phase.\cite{Aladine_PRB_2009} Instead a genuine spiral 
magnetic order is required to sustain a polar structure,\cite{Cheong_NM_2007}
with the polarity vector required to lie in the plane of the 
spiral.\cite{Cheong_NM_2007,Kundys_PRB_2009} As for the microscopic 
mechanism responsible for this interaction, the strong coupling 
between the charge and spin sectors in FeVO$_4$ has been described as a 
magnetoelastically mediated magnetostriction\cite{Kundys_PRB_2009} and as 
a magnetoelectric coupling\cite{Dixit_JPCM_2009} whose primary origin was 
proposed to lie in trilinear spin-phonon interactions.\cite{Dixit_PRB_2010}
In the magnetic sector, one of the most important terms leading to 
frustration and incommensurate ordered phases is the Dzyaloshinskii-Moriya 
interaction, which arises from spin-orbit coupling in non-inversion-symmetric 
bonding geometries. These interactions are generic in systems of low crystal 
symmetry, exactly the situation encountered in FeVO$_4$, where the triclinic 
structure has six Fe$^{3+}$ ions (three structurally inequivalent) in each 
unit cell. In combination with superexchange terms, which favor collinear 
order unless strongly frustrated, Dzyaloshinskii-Moriya interactions often 
act to produce spiral magnetic order. The resulting exchange striction, or 
lattice relaxation in the spin-ordered state, drives a polar charge state, 
i.e.~a ferroelectric. 

Finally, we comment once again that unfortunately we were not able to perform 
a direct investigation of the ferroelectric properties of FeVO$_4$ in this 
study. The weak quadrupole moment of $^{51}$V combined with the low EFG at 
the centers of the VO$_4$ tetrahedra result in a coupling between $^{51}$V 
and the crystal lattice that is too small for us to detect. Ideally, 
future studies of FeVO$_4$ would perform $^{17}$O NMR measurements on 
$^{17}$O-enriched crystals; because the O ions bridge the Fe ions and mediate 
the magnetic superexchange and super-superexchange interactions, they can be 
expected to have a much stronger quadrupolar coupling to the lattice distortion 
in the ferroelectric phase.

\section{Summary}

In summary, we have performed $^{51}$V NMR measurements on single crystals 
of FeVO$_4$ with both zero and high applied magnetic fields. We confirm 
both magnetic transitions to phases of collinear incommensurate ($T_{N1}$) 
and spiral incommensurate ($T_{N2}$) magnetic order, both occurring at values 
lower than those found for polycrystalline samples. Our data reveal a 
temperature $T^* = 65$ K marking the onset of short-ranged magnetic order 
on the NMR time scale. We observe a large Weiss constant ($\theta$) in the 
Knight shift, a prominent spectral splitting accompanying the short-range 
correlations, small magnetic moments in the ordered phases (deduced from 
the hyperfine field), and strong low-energy spin fluctuations in this regime 
(deduced from spin-lattice relaxation times). These results provide explicit 
evidence for strongly frustrated exchange interactions in FeVO$_4$, and thus 
underline the importance of magnetic frustration for the occurrence of 
improper ferroelectricity in multiferroic materials.

\acknowledgments

We thank Dr. C. Y. Wang and Prof. X. J. Zhou for x-ray (von Laue) 
measurements of the crystal alignment. Work at North China Electric Power 
University was supported by the NSF of China (Grant No.~11104070) and by 
the Scientific Research Foundation for the Returned Overseas Chinese Scholars,
State Education Ministry. Work at Renmin University of China 
was supported by the NSF of China (Grant Nos.~11374364, 11174365, and 11222433) 
and by the National Basic Research Program of China (Grant Nos.~2010CB923004, 
2011CBA00112, and 2012CB921704). Work at Fujian Institute of Research on the 
Structure of Matter was supported by the National Basic Research Program of 
China (Grant No.~2012CB921701)


\end{document}